\documentclass[]{singlecol-new}

\usepackage{natbib,stfloats}
\usepackage{mathrsfs}

\usepackage{graphics}
\usepackage{graphicx}

\theoremstyle{TH}{

}

\theoremstyle{THrm}{

}

\theoremstyle{THhit}{

}

\makeatletter
\def\theequation{\arabic{equation}}

\makeatother

\begin{document}%

\newcommand{\spd}{$sp$--$d$ }
\newcommand{\ef}{E_{\rm F}}
\newcommand{\du}{{\rm d}}
\newcommand{\e}{{\rm e}}
\newcommand{\Ang}{{\rm \AA}}

\newcommand{\be}{\begin{equation}}
\newcommand{\ee}{\end{equation}}
\newcommand{\ben}{\begin{eqnarray}}
\newcommand{\een}{\end{eqnarray}}
\newcommand{\beq}{\begin{equation}}
\newcommand{\eeq}{\end{equation}}
\newcommand{\B}{\mathrm{B}}
\newcommand{\NB}{\mathrm{NB}}
\newcommand{\RRe}{\mathrm{Re}}
\newcommand{\IIm}{\mathrm{Im}}

\setcounter{page}{1}

\LRH{G. Trambly de Laissardi\`ere and D. Mayou}

\RRH{Conductivity of Graphene with Resonant Adsorbates:
Beyond the Nearest...}

\title{Conductivity of Graphene with Resonant Adsorbates:
Beyond the Nearest Neighbor Hopping Model}

\authorA{Guy Trambly de Laissardi\`ere}

\affA{Universit\'e de Cergy-Pontoise / CNRS, \\ Laboratoire de Physique Th\'eorique et Mod\'elisation\\ F-95302 Cergy-Pontoise, France\\ \qquad E-mail: guy.trambly@u-cergy.fr}
\authorB{Didier Mayou}
\affB{Universit\'e Grenoble Alpes, Institut NEEL,\\
CNRS, Institut NEEL,\\
F-38042 Grenoble, France.\\
\qquad E-mail: didier.mayou@grenoble.cnrs.fr}

\begin{abstract}
Adsorbates on graphene can create resonances that lead to efficient electron scattering and strongly affect the electronic conductivity. Therefore a proper description of these resonances is important to get a good  insight of their effect on conductivity. The characteristics of the resonance and in particular its T-matrix depend on the adsorbate itself but also on the electronic structure of graphene. Here we show that a proper tight-binding model of graphene which includes hopping beyond the nearest-neighbor lead to sizable modifications of the scattering properties with respect to the mostly used nearest neighbor hopping model. We compare results obtained with hopping beyond the nearest-neighbor to those of our recent work Phys. Rev. Lett. 113, 146601 (2013). We conclude that the universal properties discussed in our recent work are unchanged but that a detailed comparison with experiments require a sufficiently precise tight-binding model of the graphene layer.

\end{abstract}

\KEYWORD{Graphene; Adsorbate; Conductivity; Anderson Localization; Quantum Transport Calculation}

\maketitle

\section{Introduction}

Electronic transport in graphene is sensitive in particular to local defects such as vacancies or adsorbates [1-5]. These defects are interesting in the context of functionalization which aims at controlling the electronic properties by attaching groups of atoms to graphene [6-13]. Therefore a theoretical understanding of conductivity in the presence of such defects is needed. So far the case of resonant local defects as attracted much attention because these defects strongly scatter electrons and therefore affect deeply the electronic transport properties. Yet most of the studies are done in the standard nearest neighbor hopping model of graphene. In particular in a  recent work [12] we detailed some universal regimes of transport close to the Dirac point.
The aim of the present study is to analyze the effect of hopping beyond nearest neighbors in the graphene plane. As we show the universal regime presented in the previous work [12] still exist but their domain of existence in term of electron concentration for example is affected which may be important for precise comparison with experiments.

\section{Modelisation}

We have developed a simple tight-binding (TB) scheme that reproduces ab initio electronic structure in the energy range +/- 2 eV around the Dirac energy $E_D$ [14,15]. Only $\rm p_z$ orbitals are taken into account since we are interested in electronic properties at the Fermi energy $E_F$. The Hamiltonian is,
\begin{equation}
\hat{H} = \sum_{\langle i;j \rangle} = t_{ij} \left( c_i^*c_j + c_j^*c_i \right),
\end{equation}
where the coupling element matrix $t_{ij}$ depends on the distance $r_{ij}$ between orbital $i$ and orbital $j$,
\begin{equation}
t_{ij} = - \gamma_0 \, {\rm exp} \left( q \left( 1 - \frac{r_{ij}}{a} \right) \right),
\end{equation}
with the first neighbor distance, $a = 0.142$\,nm, and the first neighbors interaction in a graphene plane, $\gamma_0 = 2.7$\,eV. The constant $q$ is fixed to have a second neighbors interaction equal to $0.1\gamma_0$ [13,10].
We consider that resonant adsobates (such as H, OH, CH$_3$...) can create a covalent bond with some carbon atoms of the graphene sheet. Then a generic model is obtained by removing the $\rm p_z$ orbital of the carbon that is just below the adsorbate [6,13].

\begin{figure}[]
\begin{center}
\includegraphics[width=8cm]{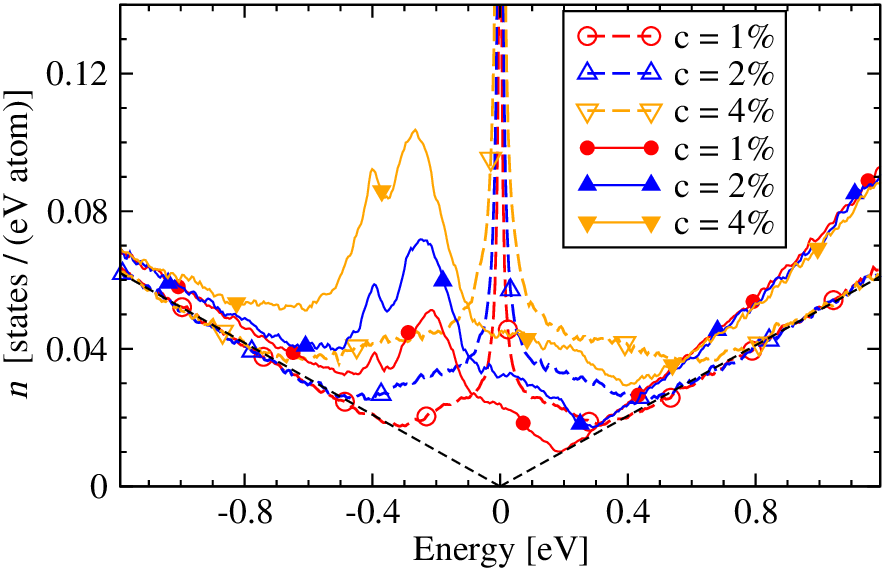}
\caption{\label{Fig1}
(colour online)
Density of states (DOS) $n$  versus energy $E$, for concentration $c$ of resonant adsorbates (vacancies):
(dashed line) first neighbor coupling only,
(solid line) beyond first neighbor coupling.
}
\end{center}
\end{figure}

In our calculations resonant adsorbates (mono-vacancies) are distributed at random with a finite concentration $c$. Preliminary results of quantum diffusion for these models have been published in [10].
The conductivity is computed by the Mayou-Khanna-Roche-Triozon (MKRT) method [16-19]. This method allows very efficient numerical calculations by recursion in real-space. Our calculations are performed on sample containing up to $10^8$ atoms. That corresponds to typical size of about one micrometer square which allows to study systems with inelastic mean-free length of the order of few hundreds nanometers. MKRT method has been also used to study other kinds of defects in graphene sheet [9,11,20,21].

In the Relaxation Time Approximation, we introduce an inelastic scattering time $\tau_i$ beyond which the propagation becomes diffusive due to the destruction of coherence by inelastic process (see Ref. [12] and Refs. therein). One finally get the conductivity,
\begin{equation}
\sigma(E_F,\tau_i) = e^2 n(E_F) D(E_F,\tau_i),
\end{equation}
and the diffusivity,
\begin{equation}
D(E_F,\tau_i) = \frac{L_i^2(E_F,\tau_i)}{2 \tau_i},
\end{equation}
where is the density of states (DOS) $n$ and $L_i$ is the inelastic mean-free path. $L_i(E,\tau_i)$ is the typical distance of propagation during the time interval $\tau_i$ for electrons at energy $E$.

\section{Results and discussion}

The total density of states for concentrations $c = 1$, 2 and 3\% are shown in figure 1. With only first neighboring coupling a resonance of the DOS is found at the Dirac energy ($E = 0$). This is reminiscent of the midgap state produced by just one missing orbital [6,13]. With coupling beyond nearest neighbors, the resonance of the density of states is enlarged and displaced by the effect of the hopping beyond nearest neighbors [13,10]. This changes of course the relation between the density of states, the electronic mean-free path $L_e$  and the Fermi energy or the filling factor (electron density) of the band.


\begin{figure}[]
\begin{center}
\includegraphics[width=8cm]{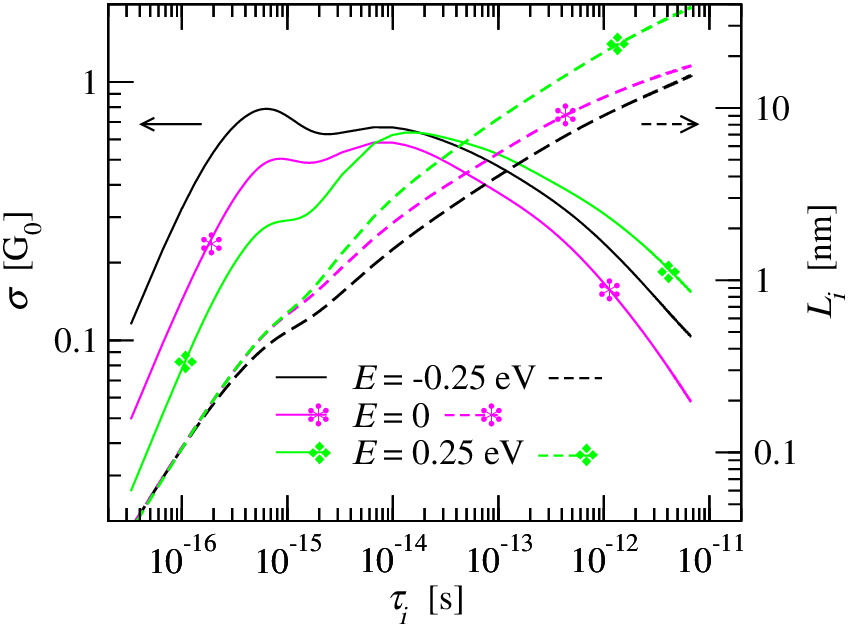}
\caption{\label{Fig2}
(colour online)
Conductivity $\sigma$ (solid line) and inelastic scattering length $L_i$ (dashed line) versus inelastic scattering time $\tau_i$ for concentration $c = 2\%$ of resonant adsorbates (vacancies), for 3 energy values and with coupling beyond first neighbor.
$G_0 = {2 e^2}/{h}$.
}
\end{center}
\end{figure}

The conductivity $\sigma$  and the inelastic scattering length $L_i$ are shown in figure 2. At small times the propagation is ballistic and the conductivity increases when $\tau_i$ increases. For large $\tau_i$, the conductivity decreases with increasing $\tau_i$ due to quantum interference effects, and it goes to zero due to Anderson localization in 2-dimension [22]. We defined the microscopic conductivity $\sigma_M$ as the maximum value of $\sigma(\tau_i)$ values (figure 2). The microscopic conductivity is shown figure 3 for different values concentration $c$. According to the renormalization theory [22] this value is obtained when the inelastic mean free path $L_i$ and the elastic mean free path $L_e$ are comparable. Here we compute $L_e$ by,
\begin{equation}
L_e(E) =\frac{2 D_M(E)}{V},
\end{equation}
where $D_M(E)$ is the maximum value of the diffusivity and $V$ is the velocity of electrons at energy $E$. $L_e$ versus $E$ is shown figure 4.
Finally we define the localization length $\xi$ as the value for which the extrapolated conductivity (see below) cancels i.e. $\sigma(L_i \simeq \xi) = 0$ (figure 5).

\begin{figure}[]
\begin{center}
\includegraphics[width=8cm]{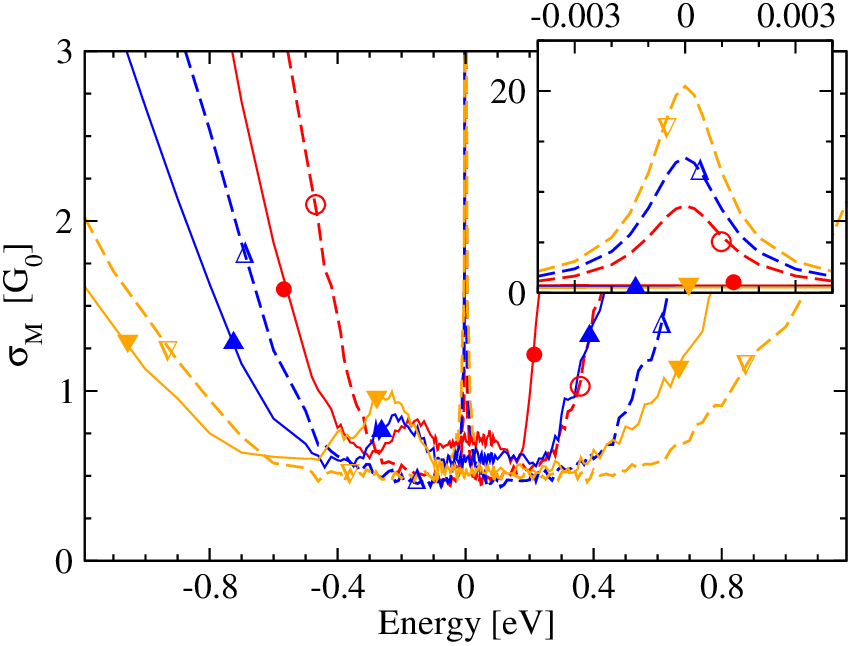}
\caption{\label{Fig3}
(colour online)
Microscopic conductivity $\sigma_M$ versus energy $E$ for concentration $c = 1$, 2 and 3\% with same symbols than figure 1: (dashed line) first neighbor coupling only, (solid line) beyond first neighbor coupling.
$G_0 = {2 e^2}/{h}$.
}
\end{center}
\end{figure}

For Fermi energy $E_F$ very close to the Dirac energy $E_D$ a strong fine peak in the microscopic conductivity is found for the model with nearest-neighbor coupling only (insert of figure 3) [23,11,12]. This particular behavior, which is a consequence of the resonance in the DOS, is specific to this model and it is not found when hopping beyond nearest neighbors is included. Indeed in the last case, the microscopic conductivity is slightly higher but still close to the universal plateau [9-12,20-21] of microscopic conductivity (figure 3), $\sigma_M \simeq 4 e^2 / (\pi h)$.
Thus our results with hopping beyond nearest neighbors show that plateau of the microscopic  conductivity near the Dirac energy is robust. On the contrary the high central peak of the conductivity and the anomalous behavior at the Dirac energy (see Ref. [12]) are not robust and are specific to the model with nearest neighbor hoping only.

For $L_e < L_i < \xi$, the conductivity is not equal to the microscopic conductivity and quantum interference have to be taken into account. As shown in figure 5, $\sigma(L_i)$ follows the linear variation with the Logarithm of the inelastic mean free path $L_i$ that was found in our previous work for smaller concentration of adsorbates [12],
\begin{equation}
\sigma(L_i) = G_0 \left(2 - \alpha\, {\rm log}\left(\frac{L_i}{L_e} \right) \right)
{\rm ~with~~}
G_0 = \frac{2 e^2}{h},
\label{equationSig_FLi}
\end{equation}
and $\alpha \simeq 0.25$, which is close to the result of the perturbation theory of 2-dimensional Anderson localization for which $\alpha = 1/\pi$ [22]. As discussed in [24,12] this could be tested through magneto-conductance measurements.

\begin{figure}[]
\begin{center}
\includegraphics[width=8cm]{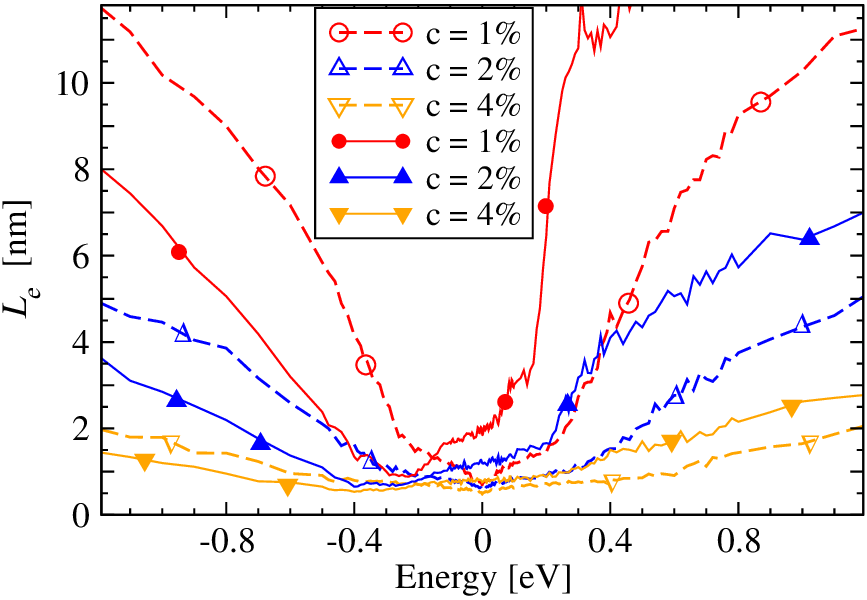}
\caption{\label{Fig4}
(colour online)
Elastic mean free path $L_e(E)$, versus energy $E$, for concentration $c$: (dashed line) first neighbor coupling only, (solid line) beyond first neighbor coupling.
}
\end{center}
\end{figure}

\begin{figure}[]
\begin{center}
\includegraphics[width=8cm]{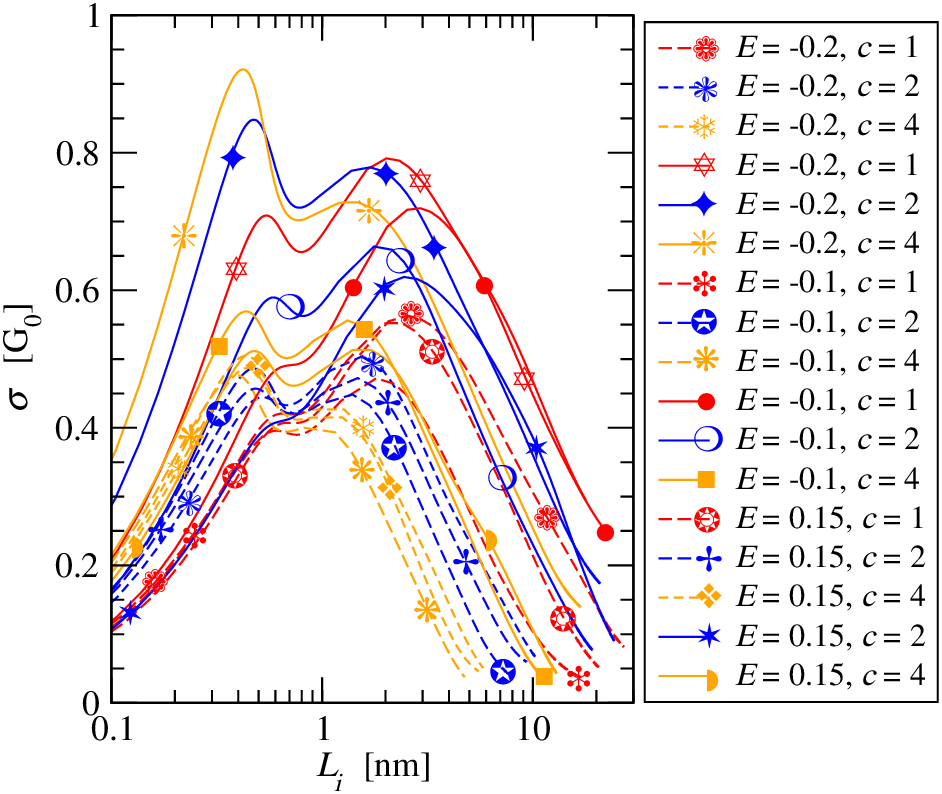}
\caption{\label{Fig5}
(colour online)
Conductivity $\sigma(L_i)$ versus the inelastic scattering length $Li$ for concentration $c$ (\%) and different energies $E$ (eV) in the plateau of $\sigma_M(E)$: (dashed line) first neighbor coupling only, (solid line) beyond first neighbor coupling.
$G_0 = {2 e^2}/{h}$.
}
\end{center}
\end{figure}

\section{Conclusion}

We propose a unified description of transport in graphene with resonant adsorbates simulated by simple vacancies[12]. Sufficiently far from the Dirac energy and at sufficiently small concentration of adsorbates the semi-classical theory is a good approximation. For Fermi energy $E_F$ close to the Dirac energy $E_D$ different quantum regimes are found.

Some universal aspects of the conductivity are present with or without the hoping beyond nearest neighbors. For small inelastic scattering length $L_i$ such as $L_i \simeq L_e$, the conductivity $\sigma$ is almost equal to the universal minimum (plateau) of microscopic conductivity $\sigma_M \simeq 4 e^2 / (\pi h)$ excepted for $E_F \simeq E_D$ when model takes only into account nearest neighbor hopping.
For larger $L_i$, $L_e < L_i$, the conductivity follows a linear variation with the Logarithm of $L_i$ [12] for both models, with nearest neighbor hopping only and with hopping beyond nearest neighbors. On the contrary the high central peak of the conductivity and the anomalous behavior at the Dirac energy (see Ref. [12]) are not robust and are specific to the model with nearest neighbor hoping only. Therefore we conclude that a precise comparison of conductivity with experiments requires a detailed description of the electronic structure and in particular of that of graphene [25].

\section*{Acknowledgment}
We thank L. Magaud, C. Berger and W. A. de Heer for fruitful discussions and comments. The computations have been performed at the Centre de Calcul of the Universit\'e de Cergy-Pontoise. We thank Y. Costes and D. Domergue for computing assistance.

\end{document}